\relax
\tolerance=1500
\magnification=1200
\baselineskip=18pt

\centerline{\bf GRAVITATIONAL SUPERENERGY TENSOR}

\

\centerline{Bahram Mashhoon$^1$, James C. McClune$^1$ and Hernando Quevedo$^2$}

\

\centerline{$^1$Department of Physics and Astronomy, University of
Missouri-Columbia}
\centerline{Columbia, MO 65211, USA}

\

\centerline{$^2$ Instituto de Ciencias Nucleares, Universidad
Nacional Aut\'onoma de M\'exico}
\centerline{Ap. Postal 70-543, 04510 M\'exico D. F., M\'exico}

\vskip .5truein

\centerline{\bf Abstract}

\

We provide a physical basis for the local gravitational superenergy
tensor. Furthermore, our gravitoelectromagnetic deduction of the
Bel-Debever-Robinson superenergy tensor permits the identification
of the gravitational stress-energy tensor. This {\it local}
gravitational analog of the Maxwell stress-energy tensor is
illustrated for a plane gravitational wave.

\


\medskip 


\vskip 0pt plus 1fill
\eject

The concept of energy and the law of conservation of energy play crucial
roles in all physical theories. The notion of particle energy 
is already present in Newtonian mechanics. In more 
general physical theories, the notion of energy becomes more
sophisticated, while still retaining its fundamental significance. In general relativity, the energy associated with matter
is represented by a stress-energy tensor $T^{\alpha\beta}_{M}$
that allows us to define, in a consistent way,
the {\it local} energy density of matter as measured by a given
observer. The condition  $T^{\alpha\beta}_{M\  ;\beta} =0$
represents the law of {\it local} conservation of energy of matter,
but it does not guarantee, in general, that the total energy is
conserved. This is because $T^{\alpha\beta}_{M}$ contains only
the contribution of matter as well as all nongravitational fields, while the gravitational energy, which
is also expected to contribute to the total energy, has not been taken into account.
However, in general relativity there is no physically meaningful notion
of energy of the gravitational field. In fact, one  would expect that in general 
relativity the gravitational energy should be given by an 
expression quadratic in the first derivatives of the metric, such as the Landau-Lifshitz pseudotensor. However, it follows from Einstein's principle of equivalence that
it is not possible to construct {\it locally} a covariant expression for energy only in
terms of the spacetime metric and its first derivatives. 
In this paper, we present a physical derivation of an expression that is,
in a certain average sense,
the {\it local} stress-energy tensor of the gravitational field, an
expression which is analogous to the stress-energy tensor of the 
electromagnetic field. In fact, the validity of this novel gravitational stress-energy tensor rests upon the approximate correspondence -- in a quasi-inertial neighborhood surrounding the worldline of a geodesic observer -- between gravitation and electromagnetism. Our gravitoelectromagnetic stress-energy tensor turns out to be directly proportional to the Bel-Debever-Robinson tensor, which is thereby furnished with a proper physical interpretation.

The electromagnetic field is endowed with a symmetric traceless
stress-energy tensor $\Theta^{\alpha \beta}$ that is given in terms
of the Faraday tensor $F^{\alpha \beta}$ by 

$$
\Theta^{\alpha \beta} = {1\over 8 \pi} (F^{\alpha \gamma}
F^\beta_{\ \gamma} +  \ ^* F^{\alpha \gamma} \ ^* F^\beta_{\ \gamma}) , 
\eqno(1)
$$

\medskip 
\noindent
where $^*F_{\alpha \beta}$ is the dual of $F_{\alpha \beta}$. In
1958, Bel suggested the notion of a gravitational superenergy tensor by constructing an analogous fourth-rank tensor
from the Riemann tensor $R_{\mu \nu \rho \sigma}$ , namely,

$$
T^{\alpha \beta \gamma \delta} = {1\over 2} (\ R^{\alpha \mu \gamma
\nu} R^{\beta\ \delta}_{\ \mu\ \nu} + \ ^*  R^{\alpha \mu \gamma \nu} \ 
^*  R^{\beta\ \delta}_{\ \mu \ \nu}\ ), \eqno(2)
$$

\medskip 
\noindent
which is unambiguously defined only for spacetimes with $R_{\mu \nu} =
\Lambda g_{\mu \nu}$ [1].  In this paper, we assume that the
cosmological constant vanishes $(\Lambda = 0)$, and the signature of
the Ricci flat spacetime is $ + 2$; moreover, units are chosen such
that Newton's constant of gravitation and the speed of light in
vacuum are set equal to unity.

The gravitational superenergy tensor (2) has been used to define the
local energy density and Poynting vector for source-free
gravitational fields [1].  Contributions to this approach have also
been made by Debever [2] and Robinson [3]; a detailed treatment of
this topic as well as further references to the original literature
is contained in Zakharov's monograph [4].

The Bel-Debever-Robinson (BDR) superenergy tensor $T^{\alpha \beta
\gamma \delta}$ is totally symmetric and traceless.  Moreover, it
satisfies a conservation law 
$T^{\alpha \beta \gamma \delta}_{\ \ \ \ \ ;\delta} = 0$ 
in analogy with $\Theta^{\alpha \beta}_{\ \ ; \beta} = 0$ for
source-free electromagnetic fields.  The BDR tensor has been particularly useful
in numerical relativity and has therefore been the subject of studies
by a number of investigators [5].  However, a basic derivation as
well as physical interpretation of the superenergy tensor has been
lacking thus far.  It is the purpose of the present work to
ameliorate this situation.

To provide a physical basis for the analogy with electrodynamics,
essential use will be made here of the notion of a {\it Fermi system}. 
 Consider an observer freely falling in a gravitational field; the
observer carries a tetrad frame $\lambda^\mu_{(\alpha)}$, where
$\lambda^\mu_{(0)} = d x^\mu/d \tau$ is the observer's velocity and
$\tau$ is the proper time along its path.  The Fermi system -- which
is the simplest generalization of a local inertial frame along the
path of the observer -- is a geodesic coordinate system based on a
nonrotating frame along the observer's worldline [6].  Let
$X^\alpha = (\tau,${\bf X}) be the Fermi coordinates along the
observer's worldline; the spacetime metric in these coordinates is
given by $^Fg_{\alpha \beta} = \eta_{\alpha \beta} +\ ^Fh_{\alpha
\beta}$, where $\eta_{\alpha \beta}$ is the Minkowski metric and $^F
h_{\alpha \beta} (\tau,${\bf X}) may be expressed as a power series
in {\bf X} away from the observer with coefficients that depend on
the Riemannian curvature of spacetime.  Only the lowest order
terms away from the path will be taken into account throughout this
paper.  It is useful to define the gravitoelectric and
gravitomagnetic potentials $\Phi_g$ and {\bf A}$_g$, respectively,
via $^F h_{00} = 2 \Phi_g$ and $^F h_{0i} = -2 (${\bf A}$_g)_i$ in
analogy with electrodynamics.

It follows from the construction of the Fermi system that 

$$
\Phi_g = - {1\over 2} \ ^F R_{0i0j} (\tau) X^i X^j , \eqno(3)
$$
\medskip
$$
( {\bf A}_g)_i = {1\over 3} \ ^F R_{0jik} (\tau) X^j X^k , \eqno(4)
$$

\medskip 
\noindent
where $^F R_{\alpha \beta \gamma \delta} (\tau) = R_{(\alpha) (\beta)
(\gamma) (\delta)} = R_{\mu \nu \rho \sigma} \lambda^\mu_{(\alpha)}
\lambda^\nu_{(\beta)} \lambda^\rho_{(\gamma)} \lambda^\sigma _{(\delta)}$ 
is the curvature as measured by the fiducial observer.  A discussion
of the general properties of this curvature is beyond the scope of
this paper [7,8].  We now define the gravitoelectric $({\cal E})$
and gravitomagnetic $({\cal  B})$ fields in complete analogy with
electrodynamics; hence, 

\medskip 

$$
{\cal E}_i (\tau,  {\bf X}) =\ ^F  R_{0i0j} (\tau) X^j , \eqno(5)
$$
\medskip 
$$
{\cal B}_i (\tau,  {\bf X}) = - {1 \over  2} \> \epsilon_{ijk}\  ^F R_{jk0l} 
(\tau) X^l , \eqno(6) 
$$

\medskip 
\noindent
which agree with the identification of ``electric" and ``magnetic"
components of the Riemann tensor.  The curvature tensor generally
consists of ``electric", ``magnetic," and ``spatial" components; in a
Ricci flat region, however, the ``spatial" components are given by
the ``electric" components.  Therefore, Eqs. (5) and (6) contain the
full information regarding the gravitational field along the path.
Moreover, it is possible to express the gravitational field equations
in the Fermi system in a form that is analogous to the Maxwell
equations using Eqs. (5) and (6).  This partial agreement between
the field equations of gravitation and electrodynamics is well known
and will not be further elaborated here [9].

The Lorentz force law is needed in addition to Maxwell's equations
to produce a complete picture of classical electrodynamics.   In the
Fermi system, this equation to first order in velocity is 
\medskip 
$$
m {d^2  {\bf X} \over d \tau^2} = q_{_E} {\cal  E} + q_{_B}
{d  {\bf X}\over d \tau} \times {\cal B}, \eqno(7)
$$

\medskip
\noindent
where $m$ is the inertial mass of a test particle, and $q_{_E}$ and
$q_{_B}$ are the gravitoelectric and gravitomagnetic charges of the
particle, respectively.  It follows from the gravitational Larmor
theorem [10] that $q_{_E} = - m$ and $q_{_B} = - 2m$; the negative signs of
the gravitoelectromagnetic charges reflect the attractive nature of
gravitation, while $q_{_B}/q_{_E} = 2$ since gravitation is a spin-$2$
field.   Equation (7) turns out to be the generalized Jacobi
equation [11], which is the equation of motion of a free test
particle {\it relative} to the fiducial observer, valid to first
order in {\bf X} and $d${\bf X}$/d \tau$.  This agreement between
the deviation equation and the Lorentz force law is a remarkable
result, since it makes it possible, in principle, to introduce the
stresss-energy tensor for gravitoelectromagnetism in complete
analogy with the standard deduction of the electromagnetic
stress-energy tensor.  However, the electromagnetic stress-energy
tensor is defined globally while the gravitoelectromagnetic
stress-energy tensor would have to be defined in a narrow tube along
the timelike path of the observer in spacetime.  The invariance
under translations in time and space is ultimately responsible for the
existence of conservation laws of energy and momentum, respectively,
as well as the stress-energy tensor in Minkowski spacetime.  The
existence of the gravitational pseudotensor can be similarly justified in
an asymptotically Minkowskian spacetime manifold.  On the other hand,
the gravitoelectromagnetic stress-energy tensor would owe its
existence to an inertial Fermi region (called the ``tidal frame" in
[12]) in the form of a thin Minkowskian cylinder in spacetime along the
timelike path of the observer.

It is possible to define a gravitational Faraday tensor in the Fermi
system by 

$$
{\cal F}_{\alpha \beta} = -\ ^F R_{\alpha \beta 0 l} X^l , \eqno(8)
$$

\medskip 
\noindent
which contains Eqs. (5) and (6).  Let us define the stress-energy
tensor associated with ${\cal F}_{ \alpha \beta}$ in the standard manner
by 

\medskip

$$
{\cal T}^{\alpha \beta} = {1\over 4 \pi} ({\cal F}^\alpha_{ \ \gamma}
{\cal F}^{\beta \gamma} - {1\over 4} \eta^{\alpha \beta} 
{\cal F}_{\gamma \delta} {\cal F}^{\gamma \delta}), \eqno(9) 
$$

\medskip
\noindent
which is symmetric and traceless by construction.  The substitution
of Eq. (8) in Eq. (9) reveals that ${\cal T}^{\alpha \beta}$ is zero
along the path of the fiducial observer ({\bf X}$= 0$); this is
ultimately a consequence of Einstein's principle of equivalence.  On
the other hand, ${\cal T}^{\alpha \beta}$ is nonzero in a neighborhood
of this path as the gravitoelectromagnetic field initially varies
linearly with distance away from the fiducial observer.  Imagine a
second observer with a worldline only infinitesimally distant from
the fiducial observer; if a second Fermi system is established along
this path and our construction of the gravitoelectromagnetic
stress-energy tensor is repeated, then this tensor would vanish
identically along the second path but would give a finite result
along the original reference path.  It follows from these
considerations that the proper stress-energy tensor along the path
of an observer must be defined through an averaging procedure.  At
any given proper time $\tau$, let us average the tensor given by Eq.
(9) over a limiting tube along the worldline.  More precisely,
consider a sphere of proper radius $\epsilon L,\ 0 < \epsilon \ll
1$, centered around {\bf X}$ = 0$ at a given $\tau$ such that

\medskip 

$$
T^{(\alpha) (\beta)} (\tau) = \lim\limits_{\epsilon \to 0} \epsilon
^{-2} < {\cal T}^{\alpha \beta} > , \eqno(10)
$$

\noindent
where the angular brackets denote the operation of averaging over the
sphere.  Here  $L$ is a constant invariant length scale that is
otherwise unspecified; in practice, it could be a characteristic
length scale for the problem under consideration (e.g., in the
exterior field of a black hole, $L$ could be the mass of the black hole), or, in the absence of a natural scale in spacetime as in a pure
gravitational radiation field, one could set $L$ equal  to the
Planck length. Using the fact that $< X^i X^j> = (\epsilon^2 L^2/3)
\delta_{ ij}$, we can express Eq. (10) in arbitrary coordinates as 

\medskip 

$$
T_{\mu \nu} = {L^2\over 12 \pi} T_{\mu \nu \rho \sigma}
\lambda^\rho_{(0)} \lambda^\sigma_{(0)} , \eqno(11) 
$$

\medskip 
\noindent
where $T_{\mu \nu \rho \sigma}$ is symmetric and traceless in the
first pair of indices and symmetric in the second pair of indices by
construction, and is given by 

\medskip

$$
T_{\mu \nu \rho \sigma} = {1\over 2} (R_{\mu \xi \rho \varsigma}
R_{\nu \ \sigma}^{\ \xi \ \ \varsigma} + R_{\mu \xi \sigma \varsigma}
R_{\nu \ \rho}^{\ \xi \ \ \varsigma}) - {1\over 4} g_{\mu \nu} R_{\alpha
\beta \rho \gamma} R^{\alpha \beta\ \gamma}_{\ \ \ \sigma} . \eqno(12) 
$$

\medskip 
\noindent
This tensor is identical with that first defined by Bel [1] in 1958
for $R_{\mu \nu} = \Lambda g_{\mu \nu}$ and coincides with Eq. (2).
It is important to remark, however, that in our
gravitoelectromagnetic deduction, i.e. Eqs. (3) - (12), no
restriction has been placed on the Ricci tensor; therefore, Eqs. (11)
and (12) define the gravitoelectromagnetic part of the local
stress-energy tensor of a general gravitational field. \
Furthermore, the linear treatment of the fields in Eqs. (5) and (6)
is sufficient to obtain the general results in Eqs. (11) and (12)
since terms of higher order would simply drop out of Eq. (10) in the limit
$\epsilon \to 0$.

To define the {\it local} gravitational stress-energy tensor at an
event in spacetime, a geodesic observer with ideal gyroscope axes is
needed at the event; then, $T_{(\alpha) (\beta)} = L^2 T_{(\alpha)
(\beta) (0) (0)}\big/ 12 \pi$ for the observer, and the 
stress-energy tensor for any other observer at that event can be obtained 
from $T_{(\alpha) (\beta)}$ by an appropriate Lorentz transformation. 
Moreover, the local physical quantities defined in this way (i.e.
energy density, energy flux, momentum density, and the gravitational
stresses) have now their proper dimensionality as a consequence of
the introduction of the constant length $L$. The freedom in the choice of $L$ implies, however, that these quantities are fixed up to a constant scale factor.

To illustrate these results, let us consider a plane gravitational
wave given by 

$$
-ds^2 = - dt^2 + dx^2 + U^2 (e^{2h} dy^2 + e^{-2h} dz^2) , \eqno(13)
$$

\noindent
which propagates along the $x$--axis and is linearly polarized.  Here
$U$ and $h$ are functions of $u = t - x$ and are related by
$U^{\prime \prime} + h^{\prime \, 2} \> U = 0$, where a prime indicates
differentiation with respect to $u$.  This gravitational field is of
Petrov type $N$ and is a member of a class of plane wave spacetimes
that has been extensively studied [13].

In the coordinate system under consideration here,  observers located
at fixed spatial coordinates follow geodesics.  Furthermore, the
natural tetrad system $\lambda^\mu_{(\alpha)}$ associated with these
observers is nonrotating.  The nonzero elements of
$\lambda^\mu_{(\alpha)}$ are given by $\lambda^0_{(0)} =
\lambda^1_{(1)} = 1,\ \lambda^2_{(2)} = U^{-1}$ exp$(-h)$, and
$\lambda^3 _{(3)} = U^{-1}$ exp$(h)$. The local gravitational
stress-energy tensor according to these observers is given by 

\medskip 

$$
(T^{(\alpha)(\beta)}) = {L^2\over 6 \pi} 
\left[ \matrix{ 1 & 1 & 0 & 0 \cr
1 & 1 & 0 & 0 \cr
0 & 0 & 0 & 0 \cr
0 & 0 & 0 & 0 \cr } \right] K^2 (u) , \eqno(14)
$$

$$
K(u) = h^{\prime \prime} + 2 {U^\prime\over U} h^\prime . \eqno(15)
$$

\noindent
The spacetime is flat if $K(u) = 0$. All the nonzero components of
$T_{(\alpha) (\beta) (\gamma) (\delta)}$ can be obtained from Eqs. (14) and (15). As expected, the only nonzero component of the
gravitational stresses is the local pressure of the radiation along
the $x$--axis.

Let $\rho_g$ be the local energy density of the gravitational
radiation field according to the fiducial static observers; then,
$\rho_g = L^2 K^2 (u)/ 6 \pi.$  What is the energy density $\hat
\rho_g$ measured by observers boosted along the $x$--axis with speed
$\beta$? The boost can be expressed in terms of coordinates as $u =
\Delta \hat u,\ \Delta v = \hat v,\ y = \hat y$, and $z
= \hat z$, where $v = t + x$ and $\Delta = (1 - \beta)^{1/2}
\big/ (1 + \beta)^{1/2}$ is the Doppler factor. By transforming the
tetrads at a given event, one can show explicitly that $\hat \rho_g =
\Delta^4 \rho_g$.  Alternatively, this result may be
obtained by noting that under the boost the metric (13) and hence the
magnitudes of $U$ and $h$ remain invariant; then, the form of Eq. (15)
leads directly to the formula for $\hat \rho_g$.

It follows that the local gravitational radiation energy density
transforms in this case like the {\it square} of the energy density
in standard field theory in Minkowski spacetime.  This could
possibly have observable consequences for the local inertia of
gravitational radiation; however, such radiation can not be
physically confined as a consequence of the universality of
gravitational interaction.

The application of these concepts to rotating gravitational waves
[14] will be published elsewhere. 

\

Two of the authors (B. M. and J. C. M.) gratefully acknowledge the
hospitality extended to them at ICN-UNAM, Mexico City, where part of
this work was done. We  wish to thank R. Sussman for helpful
discussions.

\vskip 0pt plus 1fill
\eject

\noindent
{\bf References}

\medskip

\noindent
\item{[1]} L. Bel, C. R. Acad. Sci. Paris 247 (1958) 1094;
 248 (1959) 1297; Cah. Phys. 16 (1962) 59.

\noindent
\item{[2]} R. Debever, Bull. Soc. Math. Belg. 10 (1958) 112.

\noindent
\item{[3]} See the paper of L. Bel, La Radiation Gravitationnelle, in: Colloques Internationaux du CNRS, Les Th\'eories Relativistes de la Gravitation (CNRS, Paris, 1962) pp. 119-126.

\noindent
\item{[4]} V. D. Zakharov, Gravitational Waves in Einstein's Theory,
translated by R. N. Sen from the 1972 Russian edition (Halsted, New York,
1973).

\noindent
\item{[5]} J. A. Wheeler, Phys. Rev. D 16 (1977) 3384; A. Peres,
Phys. Rev. D 18 (1978) 608; J. Ib\'a\~nez and E. Verdaguer,
Phys. Rev. D 31 (1985) 251; J. Isenberg, M. Jackson and V.
Moncrief, J. Math. Phys. 31 (1990) 517; N. Br\'eton, A.
Feinstein and J. Ib\'a\~nez, Gen. Rel. Grav. 25 (1993) 267.

\noindent
\item{[6]} For recent interesting discussions of the Fermi system see K.-P. Marzlin, Phys. Rev. D 50 (1994) 888; Gen. Rel. Grav. 26 (1994) 619.

\noindent 
\item{[7]} J. K. Beem, P. E. Ehrlich and K. L. Easley, Global Lorentzian
Geometry, second edition (Marcel Dekker, New York, 1996) ch. 2.

\noindent
\item{[8]} B. Mashhoon, Phys. Lett. A 163 (1992) 7; B. Mashhoon and
J. C. McClune, Mon. Not. Roy. Astron. Soc.  262 (1993) 881.

\noindent
\item{[9]} A. Matte, Canadian J. Math. 5 (1953) 1; B. Mashhoon, Found. Phys. 15 (1985) 497; R. T. Jantzen, P. Carini and D. Bini, Ann. Phys. (NY) 215 (1992) 1; W. B. Bonnor, Class. Quantum Grav. 12 (1995) 1483.

\noindent
\item{[10]} B. Mashhoon, Phys. Lett. A 173 (1993) 347.

\noindent
\item{[11]} B. Mashhoon, Astrophys. J. 197 (1975) 705, N.B. Eq.(A7) of the Appendix; D. E. Hodgkinson, Gen. Rel. Grav. 3 (1972) 351.

\noindent
\item{[12]} B. Mashhoon, Astrophys. J. 216 (1977) 591.

\noindent
\item{[13]} H. Bondi, F.A.E. Pirani and I. Robinson, Proc. Roy. Soc. London A
251 (1959) 519.

\noindent
\item{[14]} B. Mashhoon and H. Quevedo, Phys. Lett. A 151 (1990) 464;
H. Quevedo and B. Mashhoon, in: Relativity and Gravitation:
Classical and Quantum, Proc. SILARG VII, eds.
J. C. D'Olivo {\it et al.} (World Scientific, Singapore, 1991) pp. 287-293.

\end